\documentstyle[prl,twocolumn,aps]{revtex}
\input{epsf}
\begin{document}
\draft
\twocolumn[\hsize\textwidth\columnwidth\hsize\csname @twocolumnfalse\endcsname
\title{Breit-Wigner width and participation ratio in finite interacting Fermi
systems}

\author{B.Georgeot and D. L. Shepelyansky$^{(*)}$}

\address {Laboratoire de Physique Quantique, UMR C5626 du CNRS, 
Universit\'e Paul Sabatier, F-31062 Toulouse Cedex 4, France}

\date{22 July, 1997}

\maketitle

\begin{abstract}
For many-body Fermi systems we determine the dependence 
of the Breit-Wigner width $\Gamma$
and inverse participation ratio $\xi$ on interaction strength $U \geq U_c$
and energy excitation $\delta E \geq \delta E_{ch}$  when a crossover from
Poisson to Wigner-Dyson $P(s)$-statistics takes place.
At $U \geq U_c$ the eigenstates are composed of a large number
of noninteracting states and even for $U < U_c$ there is a regime
where $P(s)$ is close to the Poisson distribution but $\xi \gg 1$.
\end{abstract}
\vskip1pc]

\narrowtext


In 1955 Wigner \cite{wigner} introduced the local density of states
to study "the properties of the wave functions of quantum 
mechanical systems which are assumed to be so complicated that statistical 
considerations can be applied to them''. 
This quantity $\rho_W(E)$ characterizes
the spreading of eigenstates over the levels of an unperturbed system
(for example in the absence of interaction between particles), and allows
to estimate how many of these unperturbed states contribute to the real
wave function.  Generally $\rho_W(E)$ has 
a Breit-Wigner distribution with Lorentzian shape of width $\Gamma$ which
determines the energy spreading over unperturbed states.
This concept has been shown since then to be very important in a wide range of
physical problems, from nuclear physics and many-electron atoms and molecules
to condensed matter.  

The study of such complex systems has been successfully performed
through the theory of random matrices (see for example \cite{weiden}).  
Very often the physics of such systems determines some preferential basis
in which the Hamiltonian matrix has large diagonal matrix elements, while
the non-diagonal elements corresponding to transitions between the basis
states are relatively small.  The investigation of random matrices of
this type has been started only recently \cite{lenz,pichard1,dima1,jac1}.
It has been shown that the eigenstates of such superimposed band random
matrices (SBRM) are spread over the basis states
according to the Breit-Wigner distribution \cite{jac1}; this has been also
confirmed analytically through the supersymmetry approach \cite{fyodorov,frahm}.
This spreading determines the number of unperturbed states contributing
to a given eigenstate, which can be measured through the inverse participation
ratio (IPR) $\xi$ \cite{jac1,fyodorov,frahm}.
In particular the width $\Gamma$ gives an energy scale at which the
level statistics, for example the number variance $\Sigma_{2}$(E), changes 
behavior from the Wigner-Dyson to the Poisson case \cite{pichard2}.
It has been also shown that the Breit-Wigner
distribution appears in the case of sparse random matrices with preferential
basis \cite{fyodorov1}.

While the properties of the Breit-Wigner
distribution are well understood in random matrix models, the problem
of real interacting finite many-body fermionic systems was much less investigated.
Indeed, in the latter case the nature of the two-body interaction should be
taken into account, since it gives certain restrictions on the structure
of matrix elements.  A very convenient model to investigate this kind of
problem has been introduced some time ago in \cite{french,bohigas}.  This model
consists of $n$ fermions distributed over $m$ energy orbitals, coupled 
by a random two-body interaction.  Recently this model attracted a renewed
attention since it was understood that it correctly describes
the statistical properties of real physical systems such as the Ce atom and
the Si nucleus \cite{flam,zel}.  One of the main advantages of this model
is that it takes into account the two-body nature of the interaction and
allows to investigate the dependence of various quantities on the
interaction strength $U$.  This property is rather important since
the variation of the Breit-Wigner
width with respect to $U$ and excitation energy $\delta E$ counted from the
Fermi level has not been yet clearly understood.  Indeed, due to the two-body
nature of the interaction, only a small fraction of the multiparticle states
is coupled by direct transitions.  As a result, contrary to the common lore 
\cite{flam,zel,berk}, the exponential growth
of the multiparticle density of states $\rho_n$ 
with the number of particles $n$ and
the excitation energy $\delta E$ does not imply that an exponentially small
interaction leads to level mixing \cite{jac2}.  In a similar way this
exponential growth of $\rho_n$ does not lead to an exponential growth of
the width $\Gamma$.  This fact has been known in nuclear physics for some time
\cite{brody,weiden}; however, the precise dependence of $\Gamma$ on $\delta E$
has not been determined up to now.  The dependence of $\Gamma$ on $U$ is also
not obvious, due to the absence of direct coupling between the majority of 
the multiparticle states.  Different types of power-law dependence have been
recently proposed \cite{dima2,pichard3} but a definite expression for
$\Gamma$ is still elusive.  A similar situation exists for
the IPR $\xi$ in the basis of noninteracting
eigenstates which has been studied extensively very recently
\cite{altshuler,mirlin,flam2,berk2,silv}.
In this paper, we address these problems
and determine the dependence
of $\Gamma$ and $\xi$ on the parameters above.
We show that these two quantities are directly related. Surprisingly $\xi$
can be arbitrarily large at the critical interaction strength $U_c$ \cite{jac2}
where the crossover in level spacing statistics $P(s)$ 
between the Poisson and Wigner-Dyson distributions takes place.

To investigate these  properties we choose the Two-Body Random Interaction
Model (TBRIM) described above and studied recently in \cite{flam,jac2}.  In this
model $n$ fermions are located on $m$ orbitals with one-particle energies
$\epsilon_m'$ randomly chosen in the interval [$0,m$] so that the average
one-particle
level spacing is $\Delta=1$.  The multi-particle states are distributed
from the ground state $E_g \approx n^2\Delta/2$ to the maximal energy
$E_t \approx  mn\Delta-E_g$. These states are coupled by two-body random
matrix elements, varying in the interval [$-U,U$].  Due to the two-body nature
of the interaction, a given multi-particle 
state is only coupled to $K=1+n(m-n)+n(n-1)(m-n)(m-n-1)/4$
other
states in an energy interval $B=2m-4$.  
This $K$ is much smaller than
the total number of states $N=m!/n!(m-n)!$.  The density of
directly coupled states $\rho_c= K/B \approx mn^2/8$ is therefore much smaller 
than the total
density $\rho_n \approx N/(E_t-E_g)$.  
According to the results in \cite{jac2}, a crossover for $P(s)$ from Poisson to
Wigner-Dyson statistics takes place at a critical interaction strength
$U_c=C/\rho_c$ with $C \approx 0.58$.  
A similar border was also discussed in \cite{berk2}.
The precise value of $U_c$ \cite{jac2} was
determined by the condition that $\eta=\int_0^{s_0}
(P(s)-P_{WD}(s)) ds / \int_0^{s_0} (P_{P}(s)-P_{WD}(s)) ds=0.3$. 
Here $P_{P}(s)$ and $P_{WD}(s)$ are the Poisson 
and the Wigner-Dyson distributions respectively
and $s_0=0.4729...$ is their intersection point.  Physically this crossover 
happens when the coupling matrix elements become comparable to the energy
spacings between directly coupled states \cite{dima2,pichard3,jac2}.  
A similar condition determines 
the metal-insulator transition in the Anderson model, where also the level
statistics $P(s)$ changes from the Poisson distribution to the Wigner-Dyson
one \cite{shklo}.  However, the TBRIM case differs from the Anderson model
where at large system size $\eta$ can take only three values $\eta=1$
(localized), $\eta=0.215$ (critical), $\eta=0$ (delocalized), while
in the TBRIM $\eta$ varies smoothly near $U_c$ \cite{jac2}.  
Physically this difference 
comes from the fact that in the Anderson model the number of coupled neighbors
is much smaller than the linear system size, while in the TBRIM this number
is of the order of the number of states $m$ in one of $n$ directions associated
to each particle.

While the value of $U_c$ has been determined \cite{jac2}, the properties of
eigenstates as a function of the interaction remained unclear.  To understand
these properties in the TBRIM we studied the local density of states $\rho_W(E)$
in the basis of noninteracting multiparticle states.  The data were obtained 
for the states near the middle of the spectrum ($\pm  25 \% $ from the center).
The total statistics for $\rho_W$ was kept around $10^6$.
We checked that $\rho_W(E)$ has a Breit-Wigner shape and analyzed the
dependence of its width $\Gamma$ on $U$.  The numerical data for TBRIM clearly
demonstrate the relation $\Gamma \propto U^2$ which continues
up to large $U$ values where a saturation takes place (Fig. 1).  To check this
dependence for larger system sizes, we investigated a slightly different model,
obtained from the TBRIM by restricting ourselves to
states in an energy layer of width $\Delta$ near total energy $E=m\Delta$.  
Such an approximation is physically reasonable provided $\Gamma \ll \Delta$.
Indeed, in this case the transitions to states outside the layer do not
influence the properties of eigenstates.  We choose the layer to be
defined by $\sum_{i=1}^{n} m'_i=m$. The transition matrix elements
between these states were taken from the TBRIM, and the diagonal elements 
coming from one-particle energies $\epsilon_m'$ were randomly chosen in
[$(m-1/2)\Delta,(m+1/2)\Delta$]. The layer model (LM) defined in this way
retains the main physical properties of the TBRIM but allows to study systems
with much larger number of orbitals $m$.  For $n=3$ the system size of LM
is $\tilde{N} \approx m^2/12$ and for $n=4$, $\tilde{N} \approx m^3/200$.
This allowed to span $m$-values up to $m=130$ 
$(n=3, N \approx 3.6 \; 10^5)$ and $m=60$ $(n=4, N \approx 4.9 \; 10^6)$,
which are much larger than the values reached in \cite{flam,jac2,flam2}.
The multi-particle density in the LM is $\rho_n= \tilde{N} / \Delta$
while $\rho_c$ was determined numerically.
The data for LM (Fig. 1), similarly to the TBRIM case, 
also demonstrate the dependence $\Gamma \propto U^2$
and show in addition that $\Gamma \propto \rho_c$.

\begin{figure}
\epsfxsize=3.7in
\epsfysize=2.6in
\epsffile{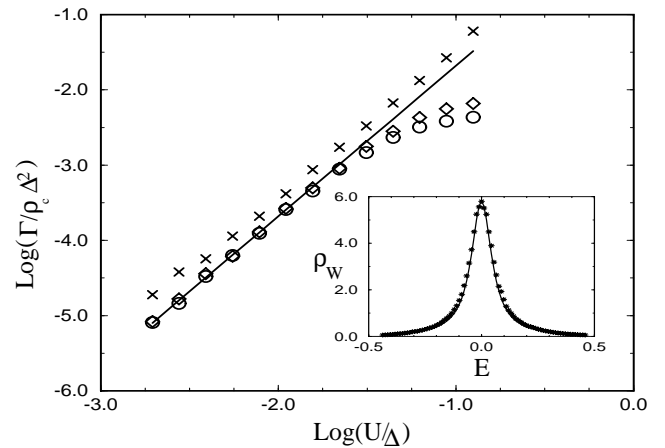}
\vglue 0.2cm
\caption{Dependence of the rescaled Breit-Wigner width $\Gamma /(\rho_c
\Delta^2)$ on $U/\Delta$: TBRIM data for $n=3,m=17$ (X); LM data for
$n=3, m=130$ (o) and $n=4, m=60$ (diamonds).  The full line shows the
theoretical estimate (1). Insert gives an example
of $\rho_W(E)$ for LM (*) with the Breit-Wigner fit $(\Gamma=0.12)$ 
for $n=3, m=130, U=0.022$ when (1) gives $\Gamma=0.125$.
} 
\label{fig1}
\end{figure}

According to the data of Fig. 1 the width $\Gamma$ is given by the Fermi golden
rule:
\begin{equation}
\label{fgr}
\Gamma = 2 \pi <U^2> \rho_c = \frac{2 \pi}{3} U^2 \rho_c.
\end{equation}
where $< ...>$ means the averaging.
We attribute the small difference between the LM and TBRIM cases
to the fact that in the latter the density $\rho_c$ slightly depends
on the energy counted from the Fermi level, while we used its average value.
For the LM this variation is smaller  and therefore the agreement is better.
 
The expression (\ref{fgr}) for $\Gamma$ doesn't depend on the multi-particle
density of states $\rho_n$.  However, for $U > U_c$ the levels are mixed
on a scale $\Delta_n=1/\rho_n \ll 1/\rho_c$ . Therefore we expect that
an eigenstate is spread over all unperturbed states in the energy interval
$\Gamma$; for $U > U_c$ this leads to the IPR:

\begin{equation}
\label{xi}
\xi \approx \Gamma \rho_n \approx 2 U^2 \rho_c \rho_n.
\end{equation} 
The numerical factor was taken in analogy with SBRM case where $\xi \approx
\Gamma \rho$ \cite{fyodorov,frahm}.
To check this theoretical estimate we computed $\xi$ for both TBRIM and LM.
The numerical data displayed in Fig. 2 show clearly the $U^2$ dependence
for sufficiently strong $U$.  At very large $U$ the growth of $\xi$ is replaced
by a saturation due to the finite size of the system.  The data shown in Fig.3 
demonstrate that $\xi \propto \rho_c \rho_n$, in agreement with (2).
\begin{figure}
\epsfxsize=3.7in
\epsfysize=2.6in
\epsffile{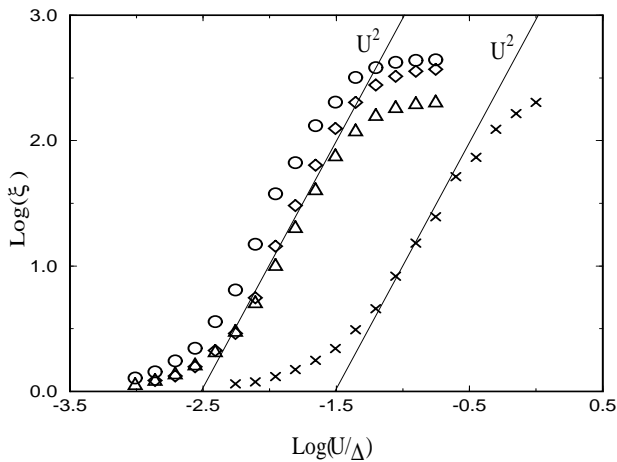}
\vglue 0.2cm
\caption{Dependence of the IPR $\xi$  on $U/\Delta$:
TBRIM data for $n=3,m=17$ (X); LM data for
$n=3, m=130$ (o), $n=3, m=90$ (triangles) and $n=4, m=60$ (diamonds).
Straight lines show dependence $\xi \propto U^2$.
} 
\label{fig2}
\end{figure}
Without any fitting parameters 
these numerical results definitely confirm the estimate (2) for
$U> U_c$.  It is interesting to check if it remains valid close
to the critical value $U_c$.  If so, then the IPR
at $U_c$ for $n \gg 1$ will contain exponentially many states 
$\xi_c = \xi (U=U_c) \sim
\rho_n/\rho_c \gg 1$.  We studied this behavior in both TBRIM and LM.
In the latter case, we checked that $U_c$ defined by the condition
$\eta(U_c)=0.3$ also follows the relation $U_c=C/\rho_c$, with $C=0.62$ being
very close to the TBRIM value \cite{jac2} (see insert in Fig. 3).  
This fact once more
confirms that indeed the LM retains the physical properties of TBRIM.  
The data of Fig. 4 for $\xi_c$ 
confirm that $\xi_c \sim \rho_n/\rho_c$ in both TBRIM and LM. In LM
the proportionality factor $\tilde{C}$ is about 3 times
smaller ($\tilde{C} \approx \rho_c \xi_c / \rho_n \approx 0.25$ in Fig. 4)
than its value given by (2) at $U=U_c$ ($\tilde{C} \approx 0.8$).
This indicates a change of eigenstate properties near $U_c$. The difference
of $\tilde{C}$ values for LM and TBRIM should be attributed to a stronger
variation of the densities $\rho_n, \rho_c$ with energy in the TBRIM.
This variation was not taken into account in the expressions for $\rho_n,
\rho_c$ in the TBRIM where we used their averaged values.
\begin{figure}
\epsfxsize=3.7in
\epsfysize=2.6in
\epsffile{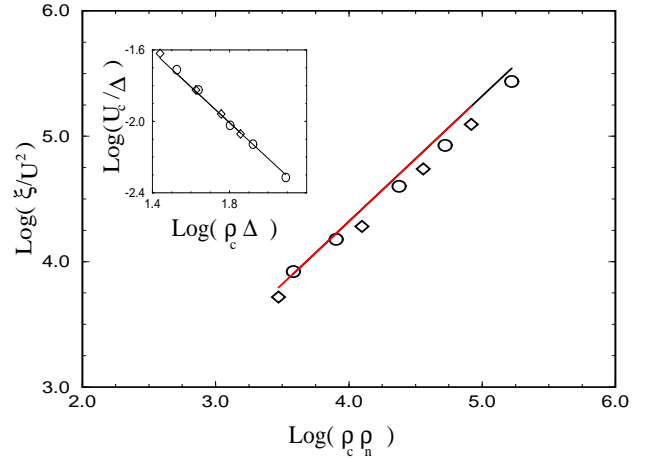}
\vglue 0.2cm
\caption{Dependence of the rescaled IPR $\xi /U^2$  on 
$\rho_c \rho_n$: LM data for $n=3$ and $40\leq m \leq 130$ (o);
$n=4$ and $30\leq m \leq 60$ (diamonds).  The straight line gives
theory (2).   Insert shows $ U_c/\Delta$ vs.
$\rho_c \Delta$ in log-log scale for the same parameters; 
the straight line is the fit $U_c = 0.62/\rho_c$.
} 
\label{fig3}
\end{figure}

The data of Fig. 4 definitely show that at $U=U_c$ the IPR grows proportionally
to the multi-particle density $\rho_n$ and, therefore, it is exponentially
large for large number of particles $n$. This fact leads to the apparently
surprising conclusion that for $U < U_c$ the eigenstates are composed of a
huge number of noninteracting eigenstates but the level statistics $P(s)$ is
still close to the Poisson distribution. A similar situation is known to exist
for quantum systems whose classical dynamics corresponds to the
Kolmogorov - Arnold - Moser (KAM) regime. In this case, the coupling between 
different modes strongly deforms the unperturbed tori, but doesn't destroys
the integrals of motion and the corresponding quantum numbers. Generally such
deformation gives a spreading over many unperturbed eigenstates, without
real mixing of energy levels \cite{chirikov}. The mixing and Wigner-Dyson
statistics for $P(s)$ appear only after the transition to chaos, 
which in our case 
corresponds to the situation when the physical frequency $1/\rho_c$
becomes comparable to the interaction induced interstate transition rate
$\Gamma$. A similar phenomenon takes place near the Anderson transition
\cite{shklo}. However, in this case $\xi$ becomes infinite at/above the
transition.

We have so far discussed the case of highly excited states
far from the Fermi level, where $\rho_n$ and $\rho_c$ are not very sensitive to
energy variation. This is not so near the Fermi energy $\epsilon_F \approx n
\Delta$, where the dependence on excitation energy $\delta E = E- E_g$ should
be taken into account. At temperature $T$ only $\delta n \sim T/\Delta$ 
particles effectively interact near the Fermi level so that
$\delta E \sim T \delta n \sim T^2/\Delta$. Since near $\epsilon_F$ the density
$\rho_c \sim \rho_2 (\delta n)^2 \sim (\delta n )^3/\Delta$ \cite{jac2}, we
obtain
\begin{eqnarray}
\label{fermi}
\Gamma \sim {\frac{U^2}{\Delta}}  \left({\frac{\delta E}{\Delta}}\right)^{3/2}
\sim {\frac{U^2}{\Delta}} \left({\frac{T}{\Delta}}\right)^2 \delta n \; ; \cr
\xi \sim \Gamma \rho_n \sim \left({\frac{\delta E}{\Delta}}\right)^{1/2} 
\exp\left(2\left(\frac{\pi^2 \delta E}{6\Delta}\right)^{1/2}\right)
\end{eqnarray} 
Here we used the known dependence of $\rho_n$ on $\delta E$ from
\cite{bohr} and assumed that $\delta E > \delta E _{ch} \approx \Delta
(\Delta/U)^{2/3}$ ($T > T_{ch} \approx \Delta (\Delta/U)^{1/3}$) 
so that the system is thermalized due to internal interaction 
($U>U_c$) \cite{jac2}.
The last expression for $\Gamma$ has a simple meaning. Indeed,
$\Gamma$ is the total spread width for $\delta n$ effectively interacting
particles. Therefore, the partial width $\Gamma_D \sim \Gamma/\delta n$
is the usual quasi-particle decay rate which in agreement with the theory of
Landau Fermi liquid is proportional to $T^2$. At the quantum chaos border 
$\delta E = \delta E_{ch}$,
when  the crossover to the Wigner-Dyson statistics takes place \cite{jac2},
the IPR becomes exponentially large $\xi_c \sim (T_{ch}/\Delta)
\exp(2.6 T_{ch}/\Delta)$.
\begin{figure}
\epsfxsize=3.7in
\epsfysize=2.6in
\epsffile{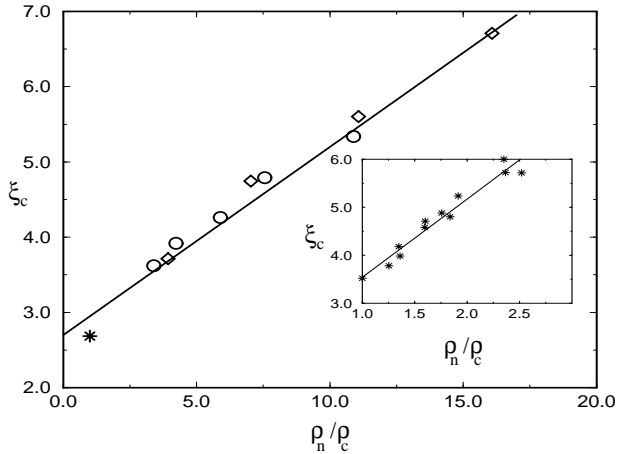}
\vglue 0.2cm
\caption{Dependence of the IPR $\xi_c$ at $U_c$  on 
$\rho_n/ \rho_c$: LM data for $n=2$ and $m=800$ (*); 
$n=3$ and $40\leq m \leq 130$ (o);
$n=4$ and $30\leq m \leq 60$ (diamonds).  The straight line gives
the fit $\xi_c= 0.25 \rho_n / \rho_c + 2.7$.   Insert shows the same
plot for the TBRIM for $n=2, m=30$; $3\leq n\leq 6$ and $10\leq m \leq 21$
(*). The straight line is the fit $\xi_c= 1.63 \rho_n / \rho_c + 1.91$.
} 
\label{fig4}
\end{figure}

In conclusion, our results allowed to understand the eigenstate
properties in finite Fermi systems
with interparticle interaction  $U \geq U_c \sim 1/\rho_c$
and excitation energy $\delta E \geq \delta E_{ch}$.
Qualitatively, these properties are similar and
opposite to the recent expectations  \cite{flam2} 
and \cite{flam,zel,berk,dima2,pichard3,altshuler,mirlin} correspondingly.
Further investigations are required for the regime with $U < U_c$
where there are indications on the appearance of another 
dependence of the IPR on system parameters \cite{berk2}.
Direct numerical checks of the relations (3) are also desirable.

We thank Ph.Jacquod for useful discussions.


\end{document}